\documentclass[useAMS,usenatbib,usegraphicx]{mn2e}
\usepackage{float}
\newcommand{\bugref}{\bibitem[\protect\citeauthoryear{dummy }{1893}]{dum}}

\title[Spine--Sheath Polarization Structures in AGN Jets]
{Are Spine--Sheath Polarization Structures in the Jets of Active 
Galactic Nuclei Associated with Helical Magnetic Fields?}

\author[D. C. Gabuzda, A. R. Reichstein \& E. L. O'Neill]
{D. C. Gabuzda, A. R. Reichstein and E. L. O'Neill \\
Physics Department, University College Cork, Cork, Ireland} 

\begin{document}

\date{}
\pagerange{\pageref{firstpage}--\pageref{lastpage}} \pubyear{2013}
\maketitle
\label{firstpage}
\begin{abstract}
One possible origin for polarization structures across jets of Active
Galactic Nuclei (AGNs) with a central ``spine'' of orthogonal magnetic
field and a ``sheath'' of longitudinal magnetic field along one or both
edges of the jet is the presence of a helical jet magnetic field.
Simultaneous Very Long Baseline Array (VLBA) polarization observations
of AGN displaying partial or full spine--sheath polarization structures were
obtained at 4.6, 5.0, 7.9, 8.9, 12.9 and 15.4 GHz, in order to search
for additional evidence for helical
jet magnetic fields, such as transverse Faraday rotation gradients
(due to the systematic change in the line-of-sight magnetic-field component
across the jet).  Results for eight sources displaying monotonic
transverse Faraday rotation gradients with significances $\geq 3\sigma$
are presented here.  Reversals in the directions of the
transverse RM gradients with distance from the core or with time are
detected in three of these AGNs. These can be interpreted as evidence
for a nested helical magnetic field structure, with different directions
for the azimuthal field component in the inner and outer regions of 
helical field. The results presented here support the idea that many
spine--sheath polarization structures reflect
the presence of helical magnetic fields being carried by these jets.
\end{abstract}

\section{Introduction}

The radio emission associated with Active Galactic Nuclei (AGNs)
is synchrotron emission, which can be linearly polarized up to about 
75\% in optically thin regions, where the polarization angle
$\chi$ is orthogonal to the projection of the magnetic field 
{\bf B} onto the plane of the sky, and 
up to 10--15\% in optically thick regions, where $\chi$ is parallel to
the projected {\bf B} (Pacholczyk 1970). Linear polarization measurements thus
provide direct information about both the degree of order and the
direction of the {\bf B} field giving rise to the observed synchrotron
radiation.  

\begin{center}
\begin{table*}
\begin{tabular}{c|c|c|c|c}
\hline
\multicolumn{5}{c}{Table 1: Source properties}\\
Source & Redshift & pc/mas & Integrated Faraday Rotation & Reference \\ 
 &  &   &  (rad/m$^2$) & \\\hline
0333+321 & 1.259 & 8.41 & $56\pm 10$ & Rusk (1988)  \\
0738+313 & 0.631 & 6.83 & $12\pm10$ & Simard--Normandin et al. (1981) \\
0836+710 & 2.218 & 8.37 & $-9\pm 3$ & Rusk (1988) \\
0923+392 & 0.695 & 7.12 & $15\pm24$ & Rusk (1988) \\
1150+812 & 1.25 & 8.40 & $107\pm8$ & Rusk (1988) \\
1633+382 & 1.813 & 8.54 & $4\pm4$ & Rusk (1988) \\
2037+511 & 1.686 & 8.56 & $-29\pm 6$ & Rusk (1988) \\
2345$-$167 & 0.576 & 6.54 & $-0.5\pm 1.1$ & Taylor et al. (2009) \\ \hline
\end{tabular}
\end{table*}
\end{center}

Multi-frequency Very Long Baseline Interferometry (VLBI) polarization
observations also provide  information about the parsec-scale 
distribution of the spectral index (optical depth) of the emitting 
regions, as well as Faraday rotation occurring between the source 
and observer.  Faraday rotation of the plane of linear  polarization 
occurs during the passage of the associated electromagnetic wave 
through a region with free electrons and a {\bf B} field with a 
non-zero component along the line of sight. When the Faraday rotation 
occurs outside the emitting region in regions of non-relativistic 
(``thermal'') plasma, the amount of rotation is given by
\begin{eqnarray}
           \chi_{obs} - \chi_o = 
\frac{e^3\lambda^{2}}{8\pi^2\epsilon_om^2c^3}\int n_{e} 
{\mathbf B}\cdot d{\mathbf l} \equiv RM\lambda^{2}
\end{eqnarray}
where $\chi_{obs}$ and $\chi_o$ are the observed and intrinsic 
polarization angles, respectively, $-e$ and $m$ are the charge and 
mass of the particles giving rise to the Faraday rotation, usually 
taken to be electrons, $c$ is the speed of light, $n_{e}$ is the 
density of the Faraday-rotating electrons, $\mathbf{B}$ is the
magnetic field, $d\mathbf{l}$ is an element along the line of 
sight, $\lambda$ is the observing wavelength, and RM (the coefficient of 
$\lambda^2$) is the Rotation Measure (e.g., Burn 1966).  Simultaneous 
multifrequency observations thus allow the determination of both 
the RM, which carries information about the electron density and 
the line-of-sight {\bf B} field in the region of Faraday rotation, 
and $\chi_o$, which carries information about the intrinsic {\bf B}-field 
geometry associated with the source projected onto the plane of the sky.

Systematic gradients in the Faraday rotation have been reported 
across the parsec-scale jets of several AGN, interpreted as 
reflecting the systematic change in the line-of-sight component of 
a toroidal or helical jet {\bf B} field across the jet (e.g.
Hovatta et al.  2012, Mahmud et al. 2013, Gabuzda et al. 2014).  
Such fields would come about in a natural way as a result 
of the ``winding up'' of an initial ``seed'' field by the rotation of 
the central accreting objects (e.g. Nakamura, Uchida \& Hirose 2001; 
Lovelace et al. 2002).

The first AGN jet found to display polarization
structure with a central ``spine'' of longitudinal polarization
(orthogonal magnetic field) with a ``sheath'' of orthogonal polarization
(longitudinal magnetic field) along the edges of the jet was 
1055+018; Attridge et al. (1999) proposed that interaction with the 
surrounding medium caused the ``sheath'' and transverse shocks propagating 
along the jet caused the ``spine''. Since the discovery of this first
example, a number of other AGN displaying this type of magnetic 
field structure have been identified (e.g. Pushkarev et al. 2005), suggesting
that the conditions giving rise to this magnetic-field structure are
not uncommon. It has also been pointed out that such polarization structure 
could reflect the presence of an overall helical magnetic field 
associated with the jet (Lyutikov et al. 2005, Pushkarev et al. 2005): 
in projection onto the plane of the sky, the predominant component of a 
helical or toroidal field associated with a roughly cylindrical length of
jet will be orthogonal near the central part of the jet and become 
longitudinal near the jet edges. 

Although it may be difficult to conclusively demonstrate which of these
pictures is correct in any individual AGN, one way to try to distinguish 
between these two scenarios is to search for additional evidence for the 
presence of helical jet magnetic fields in AGN displaying spine--sheath
or similar polarization structures, most notably transverse Faraday rotation
gradients. Another possible sign of a helical jet magnetic field is
a systematic rise in the degree of polarization toward one or both edges
of the jet, compared to the degree of polarization near the jet ridge
line. This comes about because the degree of polarization depends on
the magnetic-field component in the plane of the sky, which will be
enhanced toward the edges of a jet carrying a helical magnetic field. 


\begin{figure*}
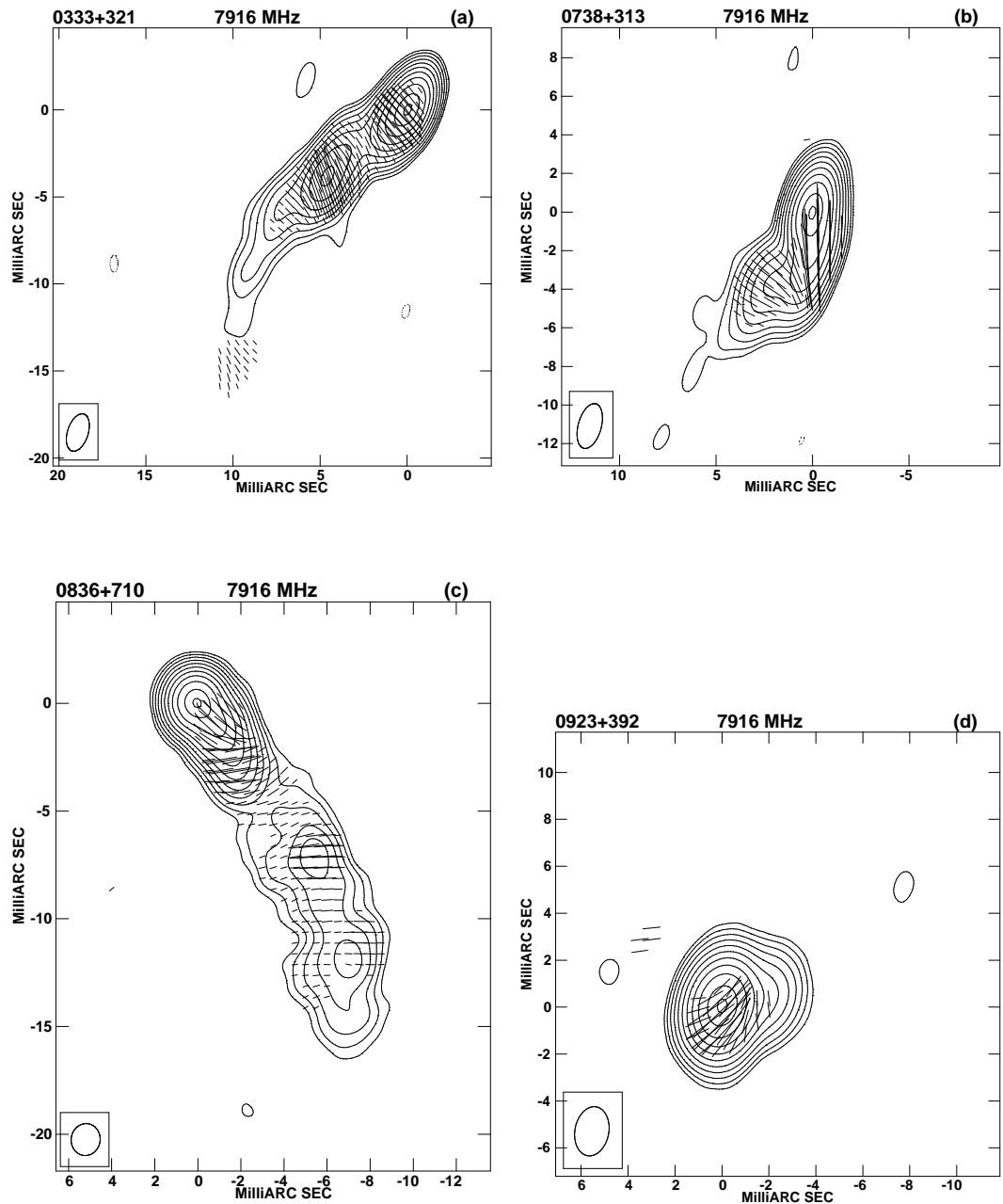

\begin{center}
\includegraphics[width=.40\textwidth,angle=0]{0333+321_POL_7.9GHZ.EPS}
\includegraphics[width=.40\textwidth,angle=0]{0738+313_POL_7.9GHZ.EPS}
\includegraphics[width=.40\textwidth,angle=0]{0836+710_POL_7.9GHZ.EPS}
\includegraphics[width=.40\textwidth,angle=0]{0923+392_POL_7.9GHZ.EPS}
\end{center}
\caption{Intensity maps with polarization angles superposed: (a) 0333+321,
peak 0.701~Jy, minimum polarized flux plotted 1.25~mJy; (b) 0738+313, 
peak 0.930~Jy, minimum polarized flux plotted 0.80~mJy; (c) 0836+710, 
peak 0.828~Jy, minimum polarized flux plotted 1.00~mJy; (d) 0923+392, 
peak 8.750~Jy, minimum polarized flux plotted 30~mJy. In all cases, the
the maps are at 7.9~GHz, and  
the contours shown are $-0.125$, 0.125, 0.25, 0.50, 1, 2, 4, 8, 16, 
32, 64 and 95\% of the peak. The beam size is shown in the lower
left corner of each map.}
\label{fig:pmaps1}
\end{figure*}

\begin{figure*}
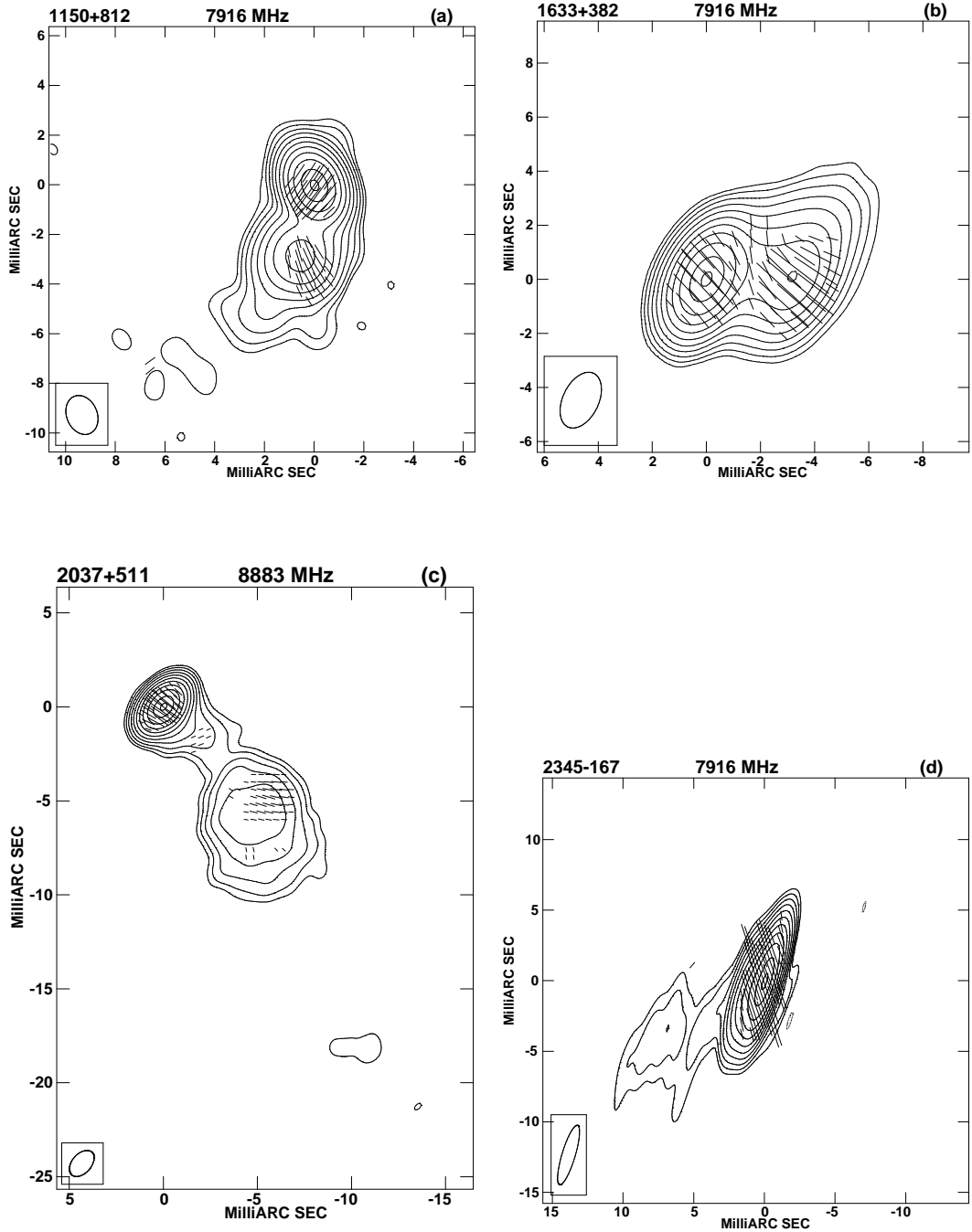

\begin{center}
\includegraphics[width=.40\textwidth,angle=0]{1150+812_POL_7.9GHZ.EPS}
\includegraphics[width=.40\textwidth,angle=0]{1633+382_POL_7.9GHZ.EPS}
\includegraphics[width=.40\textwidth,angle=0]{2037+511_POL_8.9GHZ.EPS}
\includegraphics[width=.40\textwidth,angle=0]{2345-167_POL_7.9GHZ.EPS}
\end{center}
\caption{Intensity maps with polarization angles superposed: (a) 1150+812, 
peak 1.058~Jy, minimum polarized flux plotted 3.0~mJy; (b) 1633+382, 
peak 1.517~Jy, minimum polarized flux plotted 2.8~mJy; (c) 2037+511, 
peak 1.955~Jy, minimum polarized flux plotted 1.7~mJy; (d) 2345$-$167, 
peak 1.210~Jy, minimum polarized flux plotted 2.6~mJy.  The maps of 
1150+812, 1633+382 and 
2345$-$167 are at 7.9~GHz, and the map of 2037+511
is at 8.9~GHz. In all cases, the contours shown are $-0.125$, 0.125, 0.25, 
0.50, 1, 2, 4, 8, 16, 32, 64 and 95\% of the peak. The beam size is shown 
in the lower left corner of each map.}
\label{fig:pmaps2}
\end{figure*}

With this aim, simultaneous Very Long Baseline
Array (VLBA) polarization observations of 22 AGN displaying full or
partial spine--sheath polarization structures in earlier 
MOJAVE observations ({\bf M}onitoring {\bf o}f {\bf J}ets in {\bf A}ctive
galactic nuclei with {\bf V}LBA {\bf E}xperiments;
http://www.physics.purdue.edu/MOJAVE/)
were obtained at 4.6, 5.0, 7.9, 8.9, 12.9 and 15.4 GHz, in two
24-hour sessions, on September 26 and 27, 2007 (VLBA projects BG173A
and BG173B, respectively). We consider here results for eight of these objects
observed in the latter of these two sessions.

\section{Observations and Reduction}

The observations analyzed here were obtained on September 27, 2007 
(project BG173B) with all ten antennas of the VLBA except for Saint Croix, 
which could not participate due to bad weather.  The frequencies
observed were 4.612, 5.092, 7.916, 8.883, 12.939 and 15.383 GHz. 
All of the frequencies except for 4.6 and 5.0 GHz were
built up of four intermediate frequencies (IFs) per polarization, 
each 8-MHz wide, making a total frequency bandwidth of 
32~MHz per polarization. In the case of 4.6 and 5.0~GHz, the total 
bandwidth used was split between these two frequencies, so that each 
was observed at 16-MHz per polarization.

Ten AGNs whose jets displayed signs of spine--sheath polarization 
structures in earlier MOJAVE obserations were observed in this session,
together with 0851+202, which was observed as a calibrator.
Each source was observed for a total of 25--30 minutes at each frequency, 
in a ``snap-shot'' mode with 8--10 scans spread out over the total
time the source was observable with all or most of the VLBA antennas. 
The total duration of the VLBI observations was about 24~hours. 
Eight of the ten targets observed in our September 27, 2007 session
(BG173B) showed evidence for statistically significant transverse
Faraday-rotation gradients across their core-regions and/or jets, and
it is these eight sources that we consider here (Table~1). 

The preliminary phase and 
amplitude calibration, polarization (D-term) calibration, electric
vector position angle (EVPA) calibration
and imaging were all carried out in the National Radio Astronomy
Observatory (NRAO) \textsc{AIPS} package using standard
techniques.  The reference antenna used was Los Alamos (LA).

The instrumental polarizations (D-terms) were derived using the {\sc AIPS} 
task {\sc LPCAL}, using the compact polarized source 0851+202 with 
good parallactic angle coverage, solving simultaneously for the source
polarizations in a small number of individual VLBI components corresponding
to groups of CLEAN components, identified by hand. To check the solutions, 
the D-terms for 
each of the four IFs were checked for consistency at each station, and 
plots of the real against the imaginary cross-hand polarization data were 
used to verify the successful removal of the D-terms. 

\begin{figure*}
\begin{center}
\includegraphics[width=.80\textwidth,angle=0]{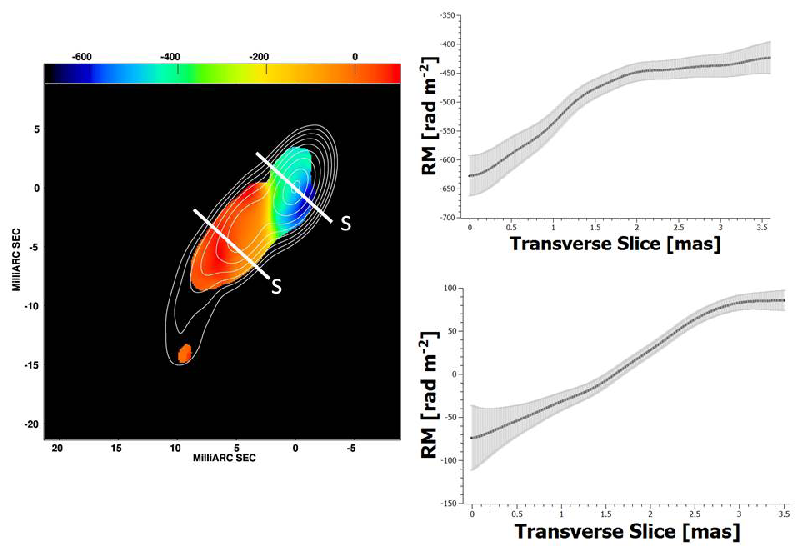}
\end{center}
\caption{4.6-GHz intensity map of 0333+321 with the RM distribution 
superposed (left).  The contours shown are $-0.25$, 0.25, 0.50, 1, 2, 4, 
8, 16, 32, 64 and 95\% of the peak of 1.05~Jy/beam. The convolving
beam is 3.8~mas $\times$ 2.0~mas in position angle $-20^{\circ}$. The lines
drawn across the RM distribution show the locations of the RM slices
shown in the corresponding right-hand panels; the letter ``S'' at one 
end of these lines marks the side corresponding to
the starting point for the slice.}
\label{fig:0333RM}
\end{figure*}

\begin{figure*}
\begin{center}
\includegraphics[width=.80\textwidth,angle=0]{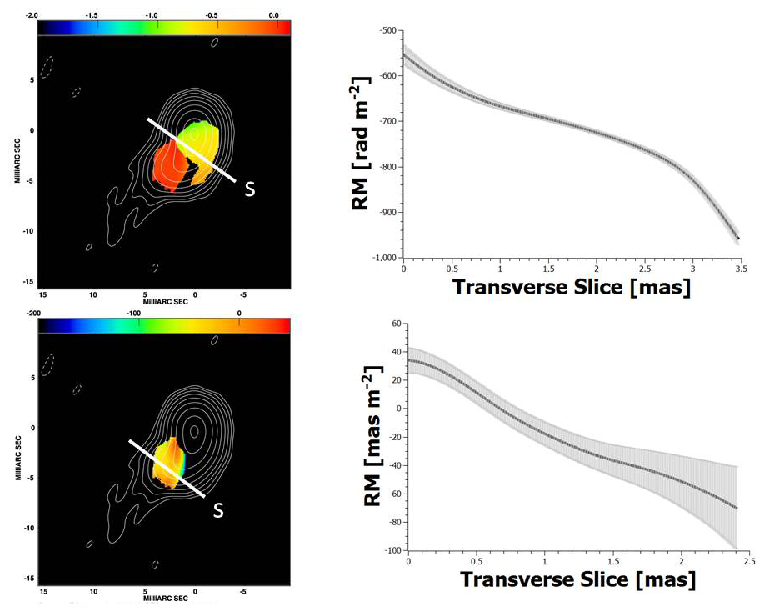}
\end{center}
\caption{4.6-GHz intensity map of 0738+318 with the RM distributions
superposed (left).  The contours shown are $-0.25$, 0.25, 0.50, 1, 2, 4, 8, 16, 
32, 64 and 95\% of the peak of 1.04~Jy/beam. The convolving
beam is 3.9~mas $\times$ 2.0~mas in position angle $-15^{\circ}$.
The colour scales for the
upper and lower RM maps have been chosen to highlight the transverse 
gradients in the core and jet regions, respectively. The lines
drawn across the RM distributions show the locations of the RM slices
shown in the corresponding right-hand panels; the letter ``S'' at one 
end of these lines marks the side corresponding to
the starting point for the slice.}
\label{fig:0738RM}
\end{figure*}

To calibrate the 
electric vector position angles (EVPAs), we used integrated polarization 
measurements of 0851+202 from the Very Large Array (VLA) polarization 
monitoring program, as well as the 26-m radio telescope of the University of 
Michigan Radio Astronomy Observatory (UMRAO). The closest VLA polarization 
observations of 0851+202 were obtained on the 29th September 2007 (5, 8.5, 
22 and 43~GHz), and the closest UMRAO observations were from the 25th 
September 2007 (14.5 GHz), 28th September 2007 (4.8 GHz) and 3rd October 
2007 (8~GHz). Fortunately, all these integrated measurements were within
a few days of the VLBA observations, minimizing the probability of
source variability between the VLBA and integrated measurements. Since 
the integrated Faraday rotation measure of 0851+202 is modest 
($+31$~rad/m$^2$; Rudnick \& Jones 1983), the differences in the
EVPA between the nearest integrated and VLBA frequencies were negligible.
The VLA and UMRAO integrated polarization angles gave consistent results. 
The rotations required for the EVPA calibration were $\Delta\chi = 
105^{\circ}, 96^{\circ}, 82^{\circ}, 87^{\circ}, 62^{\circ},$ and
$89^{\circ}$ at 4.6, 5.0, 7.9, 8.9, 12.9 and 15.4~GHz, respectively.
We estimate that the uncertainty in these EVPA calibrations is no worse
than $3^{\circ}$ at each frequency.

Maps of the distribution of the total intensity $I$ and Stokes 
parameters $Q$ and $U$ at each frequency were made, both with 
natural weighting, and with matched resolutions 
corresponding to a specified beam, usually the lowest-frequency 
beam. The distributions of the 
polarized flux ($p = \sqrt{Q^2 + U^2}$) and polarization angle 
($\chi = \frac{1}{2}\arctan \frac{U}{Q}$) were obtained from the $Q$ and $U$ 
maps using the {\sc AIPS} task {\sc COMB}. Maps of the RM were then 
constructed using the {\sc AIPS} task {\sc RM}, after first subtracting 
the effect of the integrated RM (presumed to arise in our Galaxy) from 
the observed polarization angles,  so that any residual Faraday Rotation 
was due predominantly
to thermal plasma in the vicinity of the AGN. The uncertainties in the RM
were based on the uncertainties in $Q$ and $U$, which were, in
turn, estimated using the approach of Hovatta et al. (2012). This method is
based on the results of Monte Carlo simulations, and takes into account
the fact that uncertainties on-source are higher than the off-source
rms fluctuations in the image. The output
pixels in the RM maps were blanked when the RM uncertainty from the
$\chi$ vs. $\lambda^2$ fit exceeded a specified value, usually in the 
range 50--70~rad/m$^2$.

The effect of the integrated RM is usually small, but can be 
substantial for some sources, e.g. those lying near the plane of the Galaxy. 
We used various integrated VLA RM measurements from the literature,
listed in Table~1, together with other basic information about the
eight sources for which results are presented here. The observations 
of Rusk (1988) were obtained using pointed VLA observations at 4.860 
and 14.940~GHz; we used these integrated RM values when they were 
available, which was the case for all but two of our sources. We
used the integrated RM for 2345$-$176 from Taylor et al. (2009), who
measured the Faraday Rotations of 37,543 polarized radio 
sources based on data from the NRAO VLA Sky Survey (NVSS) at
1364.9 and 1435.1 MHz. The only source for which neither of these
references provided an integrated RM was 0738+313; we used the integrated
RM for this source given by Simard--Normandin et al. (1981), who published
integrated rotation measures for 555 extragalactic radio sources based on
polarization measurements at several wavelengths between
1.73 and 10.5~GHz.  None of the
objects considered here have integrated RM values large enough to
substantially influence the observed polarization angles at the
frequencies at which our observations were made.

\section{Results}

We present here polarization maps at 7.9~GHz (8.9~GHz for 2037+511) and 
RM maps based on all six
of our frequencies for each of the eight AGNs considered in this paper.  
Additional polarization maps for each of
the sources at 15~GHz can be found in Lister \& Homan (2005) and the
MOJAVE website (http://www.physics.purdue.edu/MOJAVE/). The source names, 
redshifts, pc/mas values and integrated rotation measures are summarized 
in Table~1.  The redshifts and pc/mas vales were taken from the MOJAVE
project website (http://www.physics.purdue.edu/MOJAVE/); the latter
were determined assumed a cosmology with 
$H_o = 71$~km\,s$^{-1}$Mpc$^{-1}$, $\Omega_{\Lambda} = 0.73$ and 
$\Omega_{m} = 0.27$.

In addition to the eight AGNs considered here, we also observed two
other sources whose MOJAVE maps display signs of spine--sheath polarization 
structures: 1504$-$166 and 1730$-$130. However, the polarization and
Faraday rotation images of these sources did not reveal any new
structures, and we therefore do not consider them further here.
 
\begin{figure*}
\begin{center}
\includegraphics[width=.80\textwidth,angle=0]{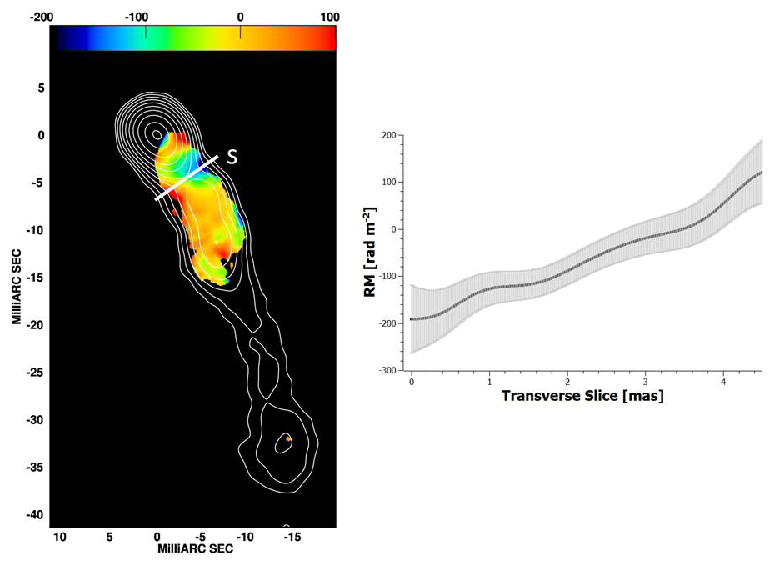}
\end{center}
\caption{4.6-GHz intensity map of 0836+710 with the RM distribution 
superposed (left).  The contours shown are $-0.25$, 0.25, 0.50, 1, 2, 4, 8, 16, 
32, 64 and 95\% of the peak of 0.98~Jy/beam. The convolving
beam is 2.8~mas $\times$ 2.8~mas. The line
drawn across the RM distribution shows the location of the RM slice
shown in the right-hand panel; the letter ``S'' at one
end of this line marks the side corresponding to
the starting point for the slice.}
\label{fig:0836RM}
\end{figure*}

\begin{figure*}
\begin{center}
\includegraphics[width=.60\textwidth,angle=0]{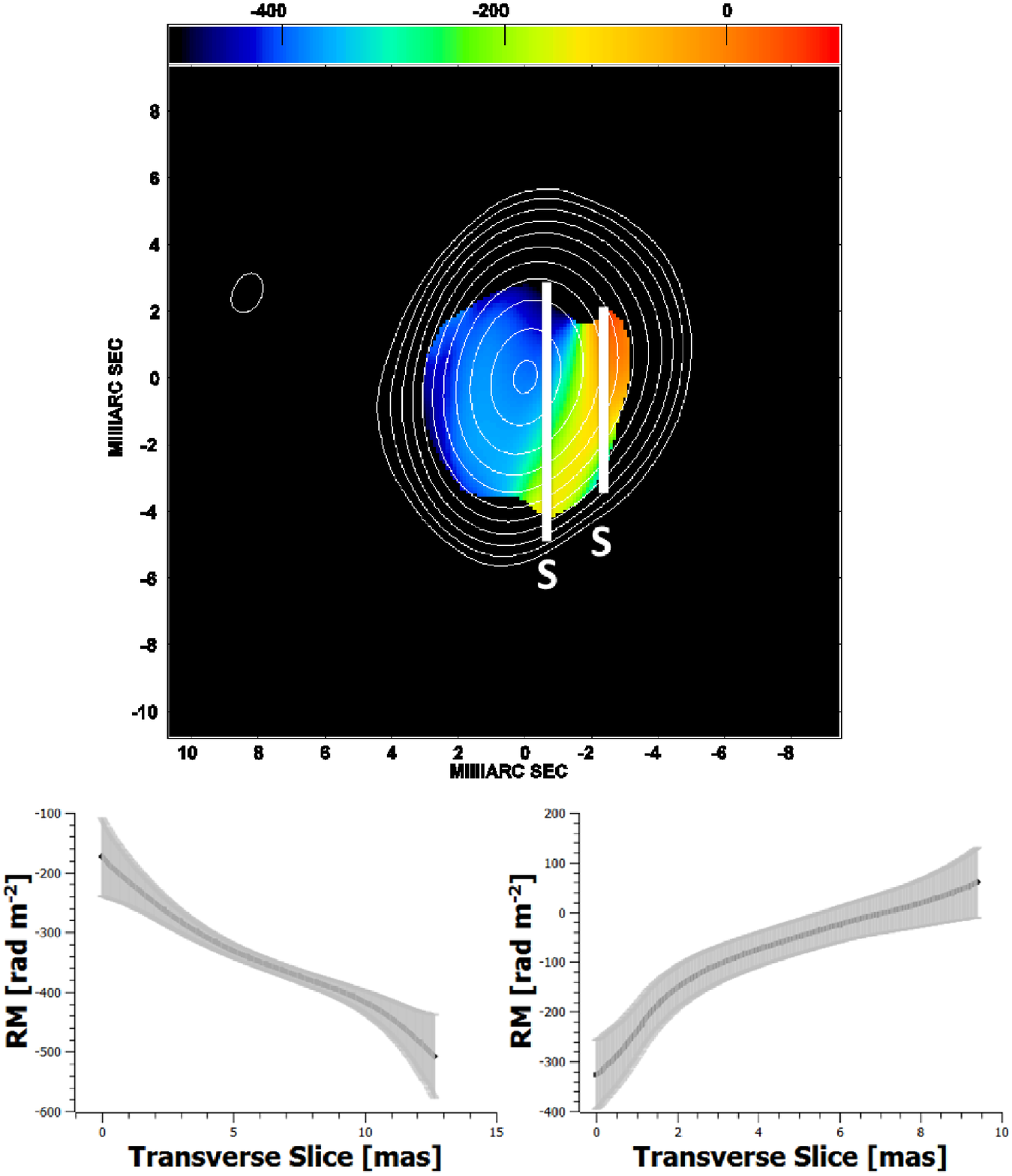}
\end{center}
\caption{4.6-GHz intensity map of 0923+392 with the RM distribution 
superposed (upper).  The contours shown are $-0.25$, 0.25, 0.50, 1, 2, 
4, 8, 16, 32, 64 and 95\% of the peak of 8.40~Jy/beam. The convolving
beam is 3.6~mas $\times$ 2.4~mas in position angle $-15^{\circ}$. The lines
drawn across the RM distributions show the locations of the RM slices
shown in the corresponding lower panels; the letter ``S'' at one
end of these lines marks the side corresponding to
the starting point for the slice.}
\label{fig:0923RM}
\end{figure*}

When possible, we determined the relative shifts between the various
images used to make the RM maps. In all cases, these shifts were tested
by making spectral-index maps taking into account the relative shifts,
to ensure that they did not show any spurious features due to residual
misalignment between the maps. The relative shifts between each frequency
and 15.4~GHz for each source for which these were significant are 
summarized in Table~2.

\begin{center}
\begin{tabular}{c|c|c|c}
\hline
\multicolumn{4}{c}{Table 2: Image shifts relative to 15.4~GHz (mas)}\\
Freq (GHz) & \multicolumn{3}{c}{Source}  \\ \hline
           & 0333+321 & 0836+710 & 2037+511\\
12.9 & 0.0 & 0.2 & 0.0\\
8.9  & 0.1 & 0.5 & 0.04\\
7.9  & 0.1& 0.7 & 0.14\\
5.0  & 0.36   & 1.2 & 0.14\\
4.6  & 0.36  & 1.4 & 0.14\\ \hline
\end{tabular}
\end{center}

RM maps together with slices in regions of detected transverse RM
gradients are also shown in Figs.~\ref{fig:0333RM}--\ref{fig:2345RM}. 
For consistency, in each case, 
these were taken in the clockwise direction relative to the base of the 
jet (located upstream from the observed core). The lines
drawn across the RM distributions show the locations of the slices; the 
letter ``S'' at one end of these lines marks the side corresponding to
the starting point for the slice (a slice distance of 0).

The statistical significance of transverse gradients detected in our
RM maps are summarized in Table~3. When plotting the slices in 
Figs.~\ref{fig:0333RM}--\ref{fig:2345RM} and finding the difference between
the RM values at two ends of a gradient, we did not include uncertainty 
in the polarization angles due to EVPA calibration uncertainty; this 
is appropriate, since EVPA calibration uncertainty affects all 
polarization angles for the frequency in question equally, and so 
cannot introduce spurious RM gradients, as is discussed by Mahmud 
et al. (2009) and Hovatta et al.  (2012).

\subsection{0333+321}

VLBA observations of 0333+321 at 8 frequencies near 4 and 8~GHz 
obtained in 2003 are presented by Asada et al. (2008),
who found the jet EVPAs to be almost perpendicular to the jet direction 
and to follow the bending of the jet. Thus, they inferred the magnetic 
field to run roughly parallel to the jet direction. Asada et al. (2008) 
also reported a clear gradient in the RM distribution across the jet, 
which they interpreted as evidence for a helical jet magnetic field, 
particularly since the gradient encompassed both positive and negative 
values.

Figure~\ref{fig:pmaps1}a shows our 7.9~GHz polarization map for 0333+321.  
As observed by Asada et al. (2008), the EVPAs are generally 
perpendicular to the jet, in both the core and jet. 
The jet polarization is offset somewhat toward the southern edge of the jet. 

We were able to derive reasonable relative core shifts for this 
source. The derived shifts relative to 15.4~GHz are given in 
Table~2. 

Figure~\ref{fig:0333RM} presents the RM distribution for 0333+321 for
our six frequencies, superimposed onto the 4.6~GHz intensity 
contours. We subtracted the integrated RM determined by Rusk (1988)
before making the RM map (Table~1); this had only a modest effect on the
observed angles, changing them by no more than about $12^{\circ}$.
The RM maps produced directly and taking into account 
the small relative shifts between the frequencies were virtually 
identical.  Our RM map confirms the presence of a clear RM gradient 
transverse to the jet direction along the entire jet and core region, 
as was reported by Asada et al. (2008).  Figure~\ref{fig:0333RM} 
also shows two slices taken roughly perpendicular to the jet 
direction plotted together with their errors.  The transverse RM 
gradients are all in the same direction and are monotonic. 
The significance of the gradients is approximately
$4\sigma$.

Inspection of the 
RM distribution by eye gives the impression of the gradient across the jet 
being slanted, as is also visible in the RM map shown in Asada et al. 
(2008). This could be due to a combination of a transverse 
RM gradient across and a decreasing RM along the jet, as would be
expected if there is a decrease in electron density and magnetic 
field strength with distance from the core. 

\subsection{0738+318}

Figure~\ref{fig:pmaps1}b shows our 7.9~GHz polarization map for 0738+318.  
The EVPAs in the inner part of the jet lie along the jet, which is
oriented toward the south, and become roughly transverse to the jet beyond
a bend toward the east a few mas from the core. The polarization beyond
this bend is offset toward the northern edge of the jet. 

\begin{figure*}
\begin{center}
\includegraphics[width=.80\textwidth,angle=0]{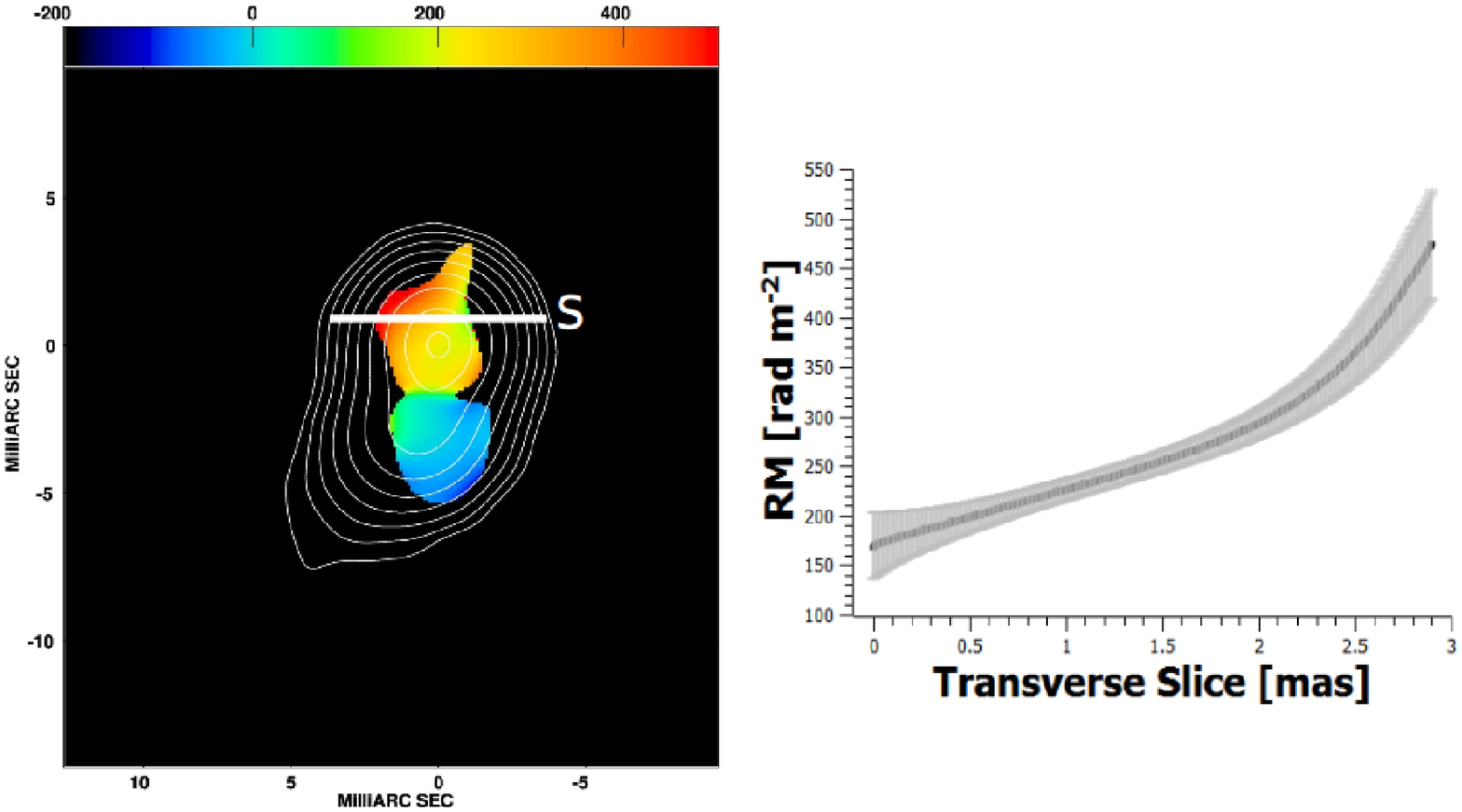}
\end{center}
\caption{4.6-GHz intensity map of 1150+812 with the RM distribution 
superposed (left).  The contours shown are $-0.25$, 0.25, 0.50, 1, 2, 4, 8, 16, 
32, 64 and 95\% of the peak of 1.08~Jy/beam. The convolving
beam is 2.5~mas $\times$ 2.1~mas in position angle $25^{\circ}$. The line
drawn across the RM distribution shows the location of the RM slice
shown in the right-hand panel; the letter ``S'' at one
end of this line marks the side corresponding to
the starting point for the slice.}
\label{fig:1150RM}
\end{figure*}

\begin{figure*}
\begin{center}
\includegraphics[width=.96\textwidth,angle=0]{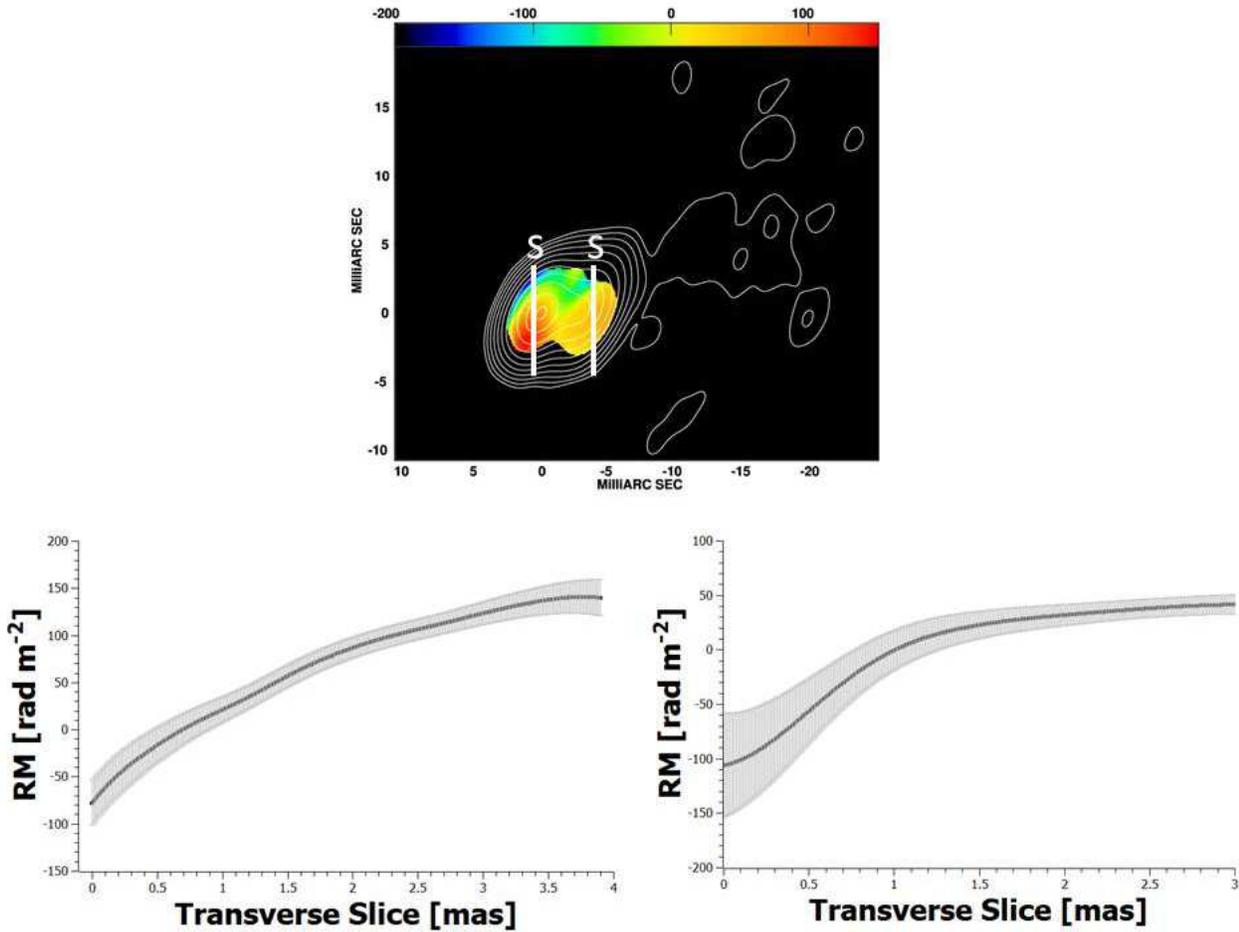}
\end{center}
\caption{4.6-GHz intensity map of 1633+382 with the RM distribution 
superposed (upper).  The contours shown are $-0.25$, 0.25, 0.50, 1, 2, 
4, 8, 16, 32, 64 and 95\% of the peak of 1.30~Jy/beam. The convolving
beam is 3.7~mas $\times$ 2.2~mas in position angle $-25^{\circ}$. The lines
drawn across the RM distributions show the locations of the RM slices
shown in the corresponding lower panels; the letter ``S'' at one
end of these lines marks the side corresponding to
the starting point for the slice.}
\label{fig:1633RM}
\end{figure*}

We found no indication of appreciable shifts between the
images at different frequencies. A spectral-index map 
produced by directly superposing the 5 and 15~GHz
images does not show any significant signs of misalignment.
We accordingly made the RM map for this source without attempting
to apply any further image alignment.

Figure~\ref{fig:0738RM} presents the RM distribution for 0738+318 for
our six frequencies, superimposed onto the 4.6~GHz intensity 
contours. We subtracted the effect of the integrated RM given
by Simard--Normandin et al. (1981) (Table~1) from the observed 
polarization angles before making the
RM map for completeness, although this had a nearly negligible
effect, changing the observed angles by no more than about $2.5^{\circ}$.
This RM map shows the presence of an RM gradient roughly
transverse to the jet direction in the core/inner jet region.
Figure~\ref{fig:0738RM} shows two slices taken roughly perpendicular to the jet 
direction plotted together with their errors.  In both
cases, the transverse RM gradients are in the same direction and 
monotonic.  The significance of the transverse RM gradient in
the inner of the two regions considered is approximately
$15\sigma$, dropping to $3.5\sigma$ further from the core. 
 
\subsection{0836+710}

VLBA observations of 0836+710 at 8 frequencies near 5 and 8~GHz 
obtained in 2003 are presented by Asada et al. (2010),
who found the projected magnetic field to be roughly 
parallel to the jet direction. They also detected a clear transverse 
RM gradient about 3--4~mas from the core.

Figure~\ref{fig:pmaps1}c shows our 7.9~GHz polarization map for 0836+710.  
As observed previously, the jet EVPAs are primarily 
perpendicular to the jet, indicating a predominantly longitudinal magnetic
field.  In some places, the polarization is clearly offset from
the center of the jet ridgeline; there are also some regions where
the polarization sticks are oblique to the jet direction. The EVPAs
in the innermost part of the jet are aligned with the jet, indicating
an orthogonal magnetic field component in this region.

We were able to derive reasonable relative core shifts for this 
source. The derived shifts relative to 15.4~GHz are given in 
Table~2. 

Figure~\ref{fig:0836RM} presents the RM distribution for 0836+710 for
our six frequencies, superimposed onto the 4.6~GHz intensity 
contours. We subtracted the effect of the integrated RM determined 
by Rusk (1988) from the observed polarization angles before making the 
RM map (Table~1) for completeness, although this had a nearly negligible
effect, changing the observed angles by no more than about $2^{\circ}$.
The RM maps produced directly and taking into account 
the small relative shifts between the frequencies were virtually 
identical.  Our RM map shows a clear gradient transverse to the 
jet roughly 5~mas from the core.  Figure~\ref{fig:0836RM} also 
shows a slice taken roughly perpendicular to the jet 
direction plotted together with its errors. The  
transverse RM gradient in this region is monotonic, and has a
significance of about $3\sigma$.

At first, these results seem to confirm the results of Asada et al. 
(2010), who observed a transverse RM gradient in approximately the 
same location in the jet. 
However, the gradient reported by Asada et al. (2010) [for data
obtained in January 2003] is in the opposite direction to the 
one observed in our data [for data obtained in September 2007]. 
In fact, Mahmud et al. (2009) observed a similar flip in the 
direction of an RM gradient across the jet of 1803+784, which
they interpreted as evidence for a picture in which the change 
in the direction of the RM gradient is due to a change in 
domination between an inner and outer region of helical
magnetic field.

\subsection{0923+392}

Figure~\ref{fig:pmaps1}d shows our 7.9~GHz polarization map for 0923+392.
Note that this represents the relatively rare case when the map peak
does not correspond to the core; the core is located at the northwest
end of the observed structure, whereas the jet structure extends toward
the southeast [see the spectral-index maps presented by Zavala \&
Taylor (2003) and Hovatta et al. (2014)].
The jet EVPAs are roughly aligned with the jet near the
jet ridgeline, becoming more orthogonal or oblique near the southern
edge of the jet. 

The core shifts between our frequencies are essentially negligible
(see also Pushkarev et al. 2012), and our RM maps were derived without 
applying additional alignments to the input data.

Figure~\ref{fig:0923RM} presents the RM distribution for 0923+392 for
our six frequencies, superimposed onto the 4.6~GHz intensity
contours. We subtracted the effect of the integrated RM determined
by Rusk (1988) from the observed polarization angles before making the
RM map (Table~1) for completeness, although this had only a small effect,
changing the observed angles by no more than about $3^{\circ}$.  Our 
RM map shows a gradient transverse to the jet in the inner jet/core 
region (northwest end of structure), as well as a transverse gradient 
in the opposite direction several mas out in the jet (southeast end of
structure). A similar structure is also
visible in the RM map of 0923+392 presented by Zavala \& Taylor (2003).

Figure~\ref{fig:0923RM} also shows slices taken roughly perpendicular to 
the jet direction plotted together with their errors.  The transverse 
RM gradients are monotonic, and have significances of approximately 
$3.9\sigma$ in the inner jet/core region and $3.4\sigma$ further
out in the jet. 

\subsection{1150+812}

Figure~\ref{fig:pmaps2}a shows our 7.9~GHz polarization map for 1150+812.
The jet EVPAs are roughly perpendicular to the jet and clearly
offset toward the western edge of the jet, indicating a longitudinal 
magnetic field, while the core EVPAs are oriented roughly parallel 
to the jet direction, suggesting an orthogonal field.

We found no indication of appreciable shifts between the
images at different frequencies. A spectral-index map
produced by directly superposing the 5 and 15~GHz
images does not show any significant signs of misalignment.
We accordingly made the RM map for this source without attempting
to apply any further image alignment.

Figure~\ref{fig:1150RM} presents the RM distribution for 1150+812 for
our six frequencies, superimposed onto the 4.6~GHz intensity
contours. We subtracted the effect of the integrated RM determined
by Rusk (1988) from the observed polarization angles before making the
RM map (Table~1). 
The core region and jet both show evidence of RM gradients 
transverse to the jet, in the same direction.  Figure~\ref{fig:1150RM} 
also shows a slice taken roughly perpendicular to the jet direction 
plotted together with its errors.  The line drawn across the RM map 
indicate the location of the slice, in the core region. This 
gradient is monotonic, 
and has a significance of approximately $5\sigma$. A transverse 
gradient that is essentially, but not strictly, monotonic can also be
observed in the inner jet. 
Although our subtraction of the integrated RM value led to an appreciable 
overall shift in the observed RM values, this did not lead to a sign
change in the RM values observed at opposite ends of the observed gradient.

\subsection{1633+382}

Figure~\ref{fig:pmaps2}b shows our 7.9~GHz polarization map for 1633+382.
The source structure is quite compact. The jet EVPAs are offset toward
the southern edge of the jet, but their orientation relative to the
jet direction is not entirely clear; the jet appears to bend toward
the north near the location of this polarization, and it is possible
that the detected polarization corresponds to a longitudinal magnetic
field at the southern edge of the jet near this bend.  
The orientation of the core EVPAs relative to the
jet direction is likewise not entirely clear. 

Due to the compactness and smoothness of the jet, it was not possible to
determine reliable shifts between the maps at the different frequencies.
Thus, our RM maps were derived without applying additional alignments to
the input data.

Figure~\ref{fig:1633RM} presents the RM distribution for 1633+382 for
our six frequencies, superimposed onto the 4.6~GHz intensity
contours. We subtracted the effect of the integrated RM determined
by Rusk (1988) from the observed polarization angles before making the
RM map (Table~1) for completeness, although this had a negligible
effect, changing the observed angles by no more than about $1^{\circ}$.
This map shows a clear transverse RM gradient all
along the core and inner jet region.  Figure~\ref{fig:1633RM} also shows 
two slices taken roughly perpendicular to the jet
direction plotted together with their errors.  The
transverse RM gradients are monotonic, and have significances of 
approximately $7\sigma$ in the core region and $3\sigma$ in the jet.

\subsection{2037+511}

In nearly all cases, our 7.9~GHz and 8.9~GHz polarization maps were very 
similar; however, the 8.9~GHz polarization map for 2037+511 was somewhat
cleaner than the 7.9~GHz map (i.e., it showed more polarization), and
we accordingly show the 8.9~GHz map here in Figure~\ref{fig:pmaps2}c.
The inner jet shows orthogonal polarization (a longitudinal magnetic field)
that is well centered on the jet spine; the strongest region of polarization 
roughly 5~mas from the core lies oblique to the jet direction and is offset 
toward the western edge of the jet. The core EVPAs are well aligned with
the direction of the inner jet, suggesting an orthogonal magnetic field.

\begin{figure*}
\begin{center}
\includegraphics[width=.80\textwidth,angle=0]{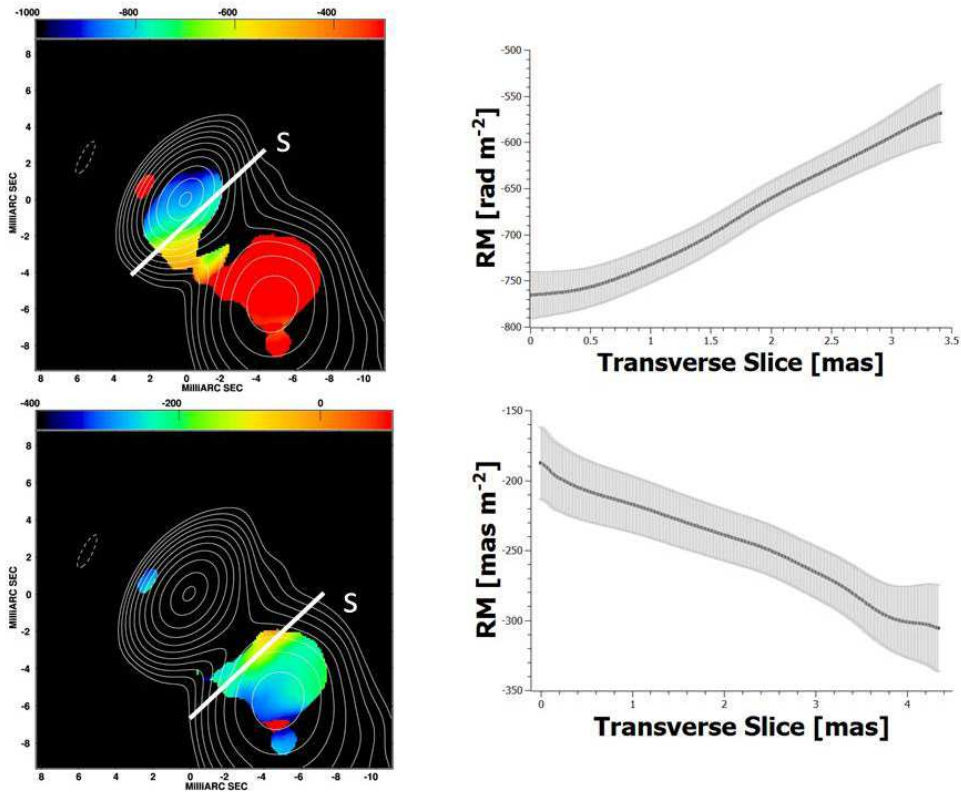}
\end{center}
\caption{4.6-GHz intensity map of 2037+511 with the RM distributions 
superposed.  The contours shown are $-0.25$, 0.25, 0.50, 1, 2, 4, 8, 16, 
32, 64 and 95\% of the peak of 1.18~Jy/beam. The convolving
beam is 3.2~mas $\times$ 1.9~mas in position angle $-40^{\circ}$. 
The colour scales for the
upper and lower RM maps have been chosen to highlight the transverse
gradients in the core and jet regions, respectively. The lines
drawn across the RM distributions show the locations of the RM slices
shown in the corresponding right-hand panels; the letter ``S'' at one
end of these lines marks the side corresponding to
the starting point for the slice.}
\label{fig:2037RM}
\end{figure*}

\begin{figure*}
\begin{center}
\includegraphics[width=.80\textwidth,angle=0]{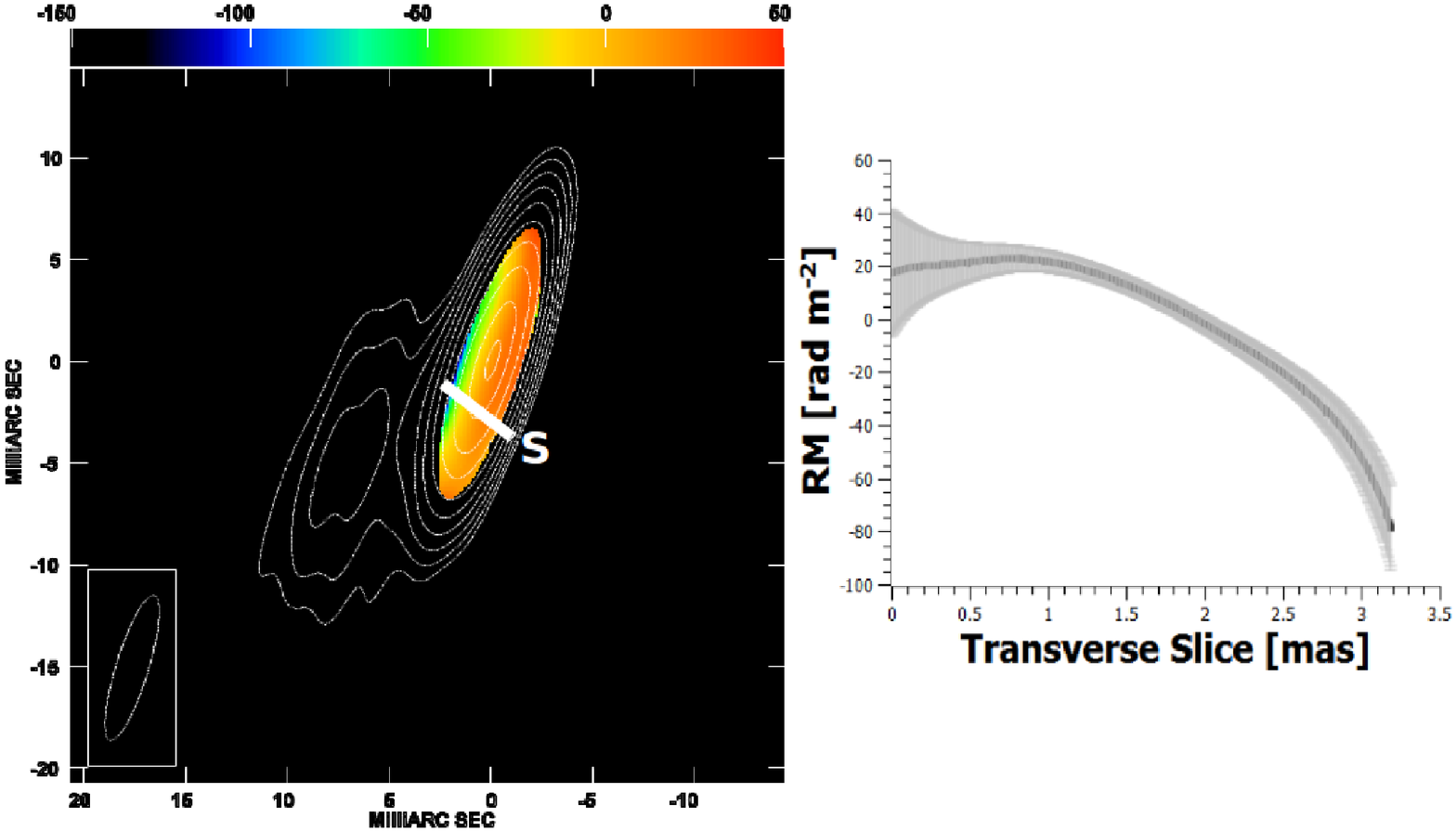}
\includegraphics[width=.80\textwidth,angle=0]{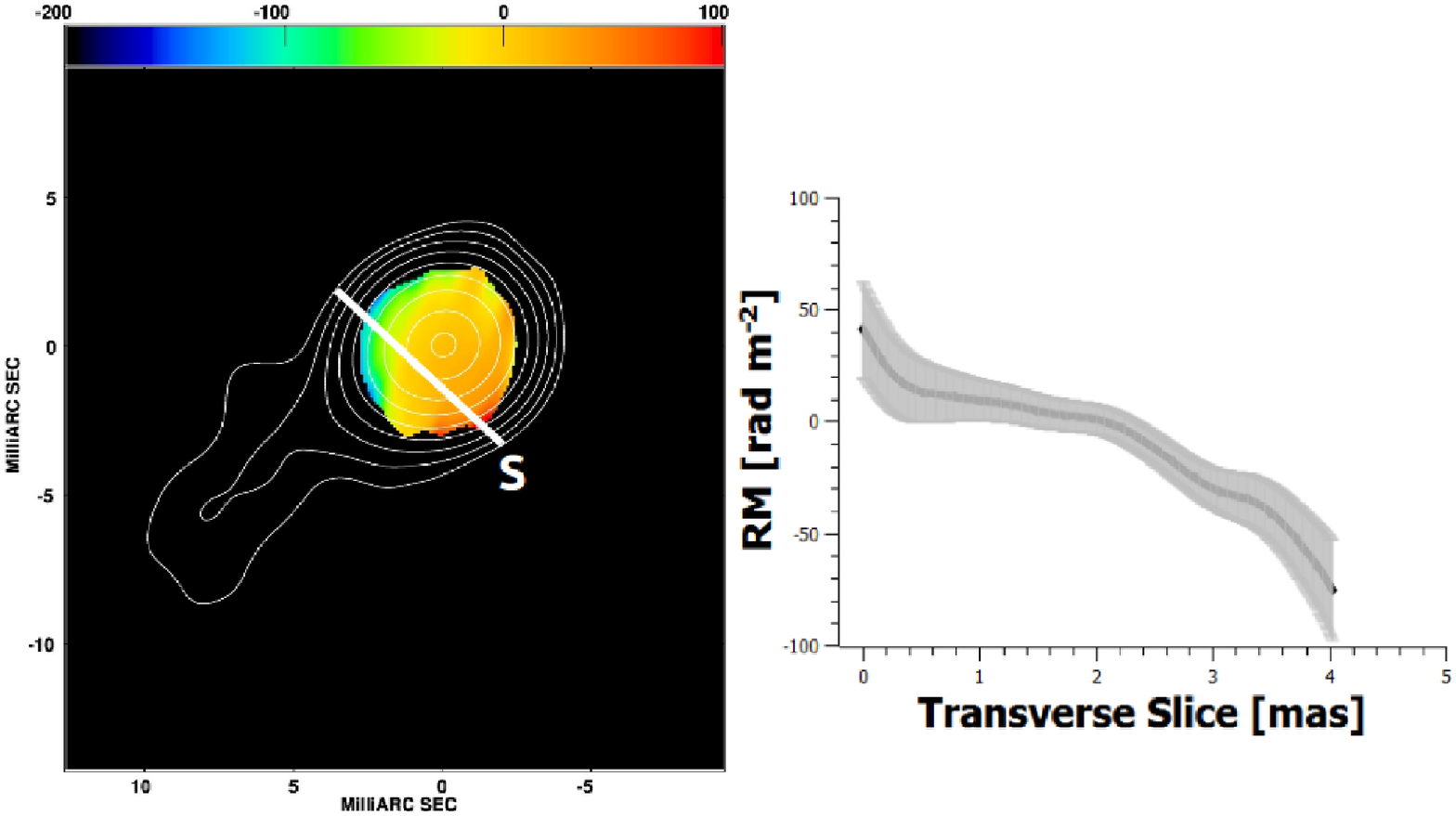}
\end{center}
\caption{4.6-GHz intensity map of 2345$-$167 with the RM distribution 
superposed, for convolution with the intrinsic elliptical beam (top left,
7.4~mas $\times$ 1.6~mas in position angle $-17^{\circ}$)
and a circular beam with the same area (bottom left).  The contours shown 
are $-0.25$, 0.25, 0.50, 1, 2, 4, 8, 16, 32, 64 and 95\% of the peaks 
of 1.54~Jy/beam (top left) and 1.61~Jy/beam (bottom left).  The lines
drawn across the RM distributions show the locations of the RM slices
shown in the corresponding right-hand panels; the letter ``S'' at one
end of these lines marks the side corresponding to
the starting point for the slice.}
\label{fig:2345RM}
\end{figure*}

We were able to derive reasonable relative core shifts for this
source, which were all quite small. The derived shifts relative to 
15.4~GHz are given in Table~2.

\begin{center}
\begin{table*}
\begin{tabular}{c|c|c|c|c|c}
\hline
\multicolumn{6}{c}{Table 3: Detected transverse RM gradients}\\
Source & Location & RM$_1$ & RM$_2$ & |$\Delta$RM| & Significance \\ 
       &          & rad/m$^2$ & rad/m$^2$ & rad/m$^2$ & \\
0333+321 & Core & $-628\pm35$ & $-424\pm 28$  & $204\pm 45$ & $4.5\sigma$\\
         & Jet  & $-73\pm35$ & $+85\pm 12$  & $158\pm 37$ & $4.3\sigma$\\
0738+313 & Core & $-553\pm22$ & $-957\pm 14$ & $404\pm 27$ & $15\sigma$\\
         & Jet  & $+34\pm9$ & $-70\pm 28$ & $104\pm 30$ & $3.5\sigma$\\
0836+710 & Jet  & $-191\pm 70$ & $+125\pm 70$ & $316\pm 100$ & $3.2\sigma$\\
0923+392 & Core & $-326\pm 70$ & $+59\pm 70$ & $386\pm 99$  & $3.9\sigma$ \\
         & Jet  & $-174\pm 69$ & $-508\pm 69$ & $334\pm 98$  & $3.4\sigma$ \\
1150+812 & Jet  & $+480\pm 45$ & $+163\pm 37$ & $317\pm 59$ & $5.3\sigma$\\
1633+382 & Core & $-78\pm 24$ & $+140\pm 20$ & $218\pm 32$ & $6.8\sigma$\\
         & Jet  & $-103\pm 45$ & $+41\pm 9$ & $144\pm 46$ & $3.1\sigma$\\
2037+511 & Core & $-765\pm 24$ & $-568\pm 25$ & $197\pm 38$ & $5.1\sigma$\\
         & Jet  & $-189\pm 30$ & $-303\pm 24$ & $114\pm 38$ & $3.0\sigma$\\
2345$-$167 -- Elliptical beam & Core & $-199 \pm 38$ & $+90\pm 31$ & $289\pm 49$ & $5.9\sigma$\\ 
2345$-$167 -- Circular beam & Core & $-58\pm 18$ & $+48\pm 21$ & $106\pm 28$ & $3.7\sigma$\\ \hline
\end{tabular}
\end{table*}
\end{center}

Figure~\ref{fig:2037RM} presents the RM distribution for 2037+511 for
our six frequencies, superimposed onto the 4.6~GHz intensity
contours. We subtracted the effect of the integrated RM determined
by Rusk (1988) from the observed polarization angles before making the
RM map (Table~1); this led to only small adjustments in the observed
polarization angles by no more than about $6^{\circ}$.
The RM maps produced directly and taking into account
the small relative shifts between the frequencies were very similar.
Our RM map shows gradients transverse to the jet in both the core and
inner jet, with opposite directions.  
Figure~\ref{fig:2037RM} shows two RM maps made with different 
RM ranges, to highlight the structure of the RM distribution in the core
region (upper) and jet (lower).  Figure~\ref{fig:2037RM} also shows two 
slices taken roughly perpendicular to the jet direction plotted together 
with their errors.  The transverse RM gradients are monotonic, and have
significances of approximately $5\sigma$ in the core region and $3\sigma$
in the jet.

\subsection{2345$-$167}

Figure~\ref{fig:pmaps2}d shows our 7.9~GHz polarization map for 2345$-$167.
The core polarization is roughly orthogonal to the jet direction, suggesting
a longitudinal magnetic field.

Due to the compactness of the source structure, it was not possible to
determine reliable shifts between the maps at the different frequencies.
Thus, our RM maps were derived without applying additional alignments to
the input data.

The top panel of Figure~\ref{fig:2345RM} presents the RM distribution for 
2345$-$167 for our six frequencies, superimposed onto the 4.6~GHz intensity
contours.  Because the integrated RM 
of Taylor et al. (2009) was very small, and would have led to completely 
negligible changes in the observed polarization angles, we did not subtract 
the effect of this integrated RM from the observed polarization angles for 
this source.  A transverse RM gradient is visible across the region just 
southeast of the core. Figure~\ref{fig:2345RM} 
also shows a slice taken roughly perpendicular to the jet direction in
this region, plotted together with its errors.  The observed transverse RM 
gradient is monotonic, and has a significance of about $6\sigma$. Although
the Monte Carlo simulations of Mahmud et al. (2013) and Murphy \& Gabuzda
(2013) have shown that reliable RM gradients can be detected even across
very narrow structures, we tested the reality of this gradient by 
convolving with a circular beam with comparable area, with a full width 
at half maximum of 2.8~mas. The resulting RM map is shown in the lower
panel of Fig.~\ref{fig:2345RM}. The transverse RM gradient is still
visible, and the slice shown indicates that it is monotonic. The statistical
significance has been reduced by the convolution with a wider beam, but
remains significant, at $3.5\sigma$.

\section{Discussion}

\subsection{Reliability and Significance of the Transverse RM Gradients}

Table~3 gives a summary of the transverse Faraday rotation measure gradients 
detected in the images presented here. The statistical significances of
these gradients typically range from 
$3\sigma$ to $5\sigma$, with one value reaching nearly $15\sigma$. This
indicates that none of the transverse RM gradients reported here are 
likely to be spurious (i.e., due to inadequacy of the $uv$ coverage
and noise): the Monte Carlo simulations of Hovatta et al. (2012) and
Murphy \& Gabuzda (2013) have shown that spurious gradients at the
$3\sigma$ level should arise for 4--6-frequency VLBA observations in the 
frequency interval considered here with a probability of no more than about
1\%, with this probability being even lower for monotonic gradients 
encompassing differences of greater than $3\sigma$. Although the occurrence
of spurious gradients rises substantially for VLBA observations of 
low-declination sources, the high significance of the gradient observed
across the inner jet of 2345$-$167 with its intrinsic (elliptical) beam -- 
about $6\sigma$ -- is high enough
to make it very improbable that this gradient is spurious.

In principle, it is optimal to ensure that the polarization angle maps
are correctly aligned relative to each other using the measured VLBI
core shifts before using them to construct RM maps. It was possible to
find reliable core shifts for three of the AGNs considered here (0333+321,
0836+710, 2037+511); in three other cases, the core shifts were estimated to 
be negligible (0738+313, 0923+392, 1150+812); in the remaining two cases, 
it was not possible to get reliable core-shift estimates due to the compactness
of the source structure (1633+382, 2345$-$167). The similarity
of the RM maps made applying and not applying the relative shifts between the
different frequencies in the three cases where this could be done 
gives us confidence that the RM maps for the three sources 
for which core shifts could not estimated are unlikely to be significantly
affected by image misalignment.

The Monte Carlo simulations of Mahmud et al. (2013) and Murphy \& Gabuzda 
(2013) have showed that it is not necessary or meaningful to place a limit 
on the width
spanned by an RM gradient in order for it to be reliable, as long as the
gradient is monotonic and the difference between the RM values at either 
end is at least $3\sigma$. 
We have used the single-pixel error estimates 
recommended by Hovatta et al. (2012) based on their own Monte Carlo 
simulations, to ensure that our single-pixel RM uncertainties are not 
underestimated. 

Although the signal-to-noise ratio for our RM estimates could in principle
be increased by averaging together some number of pixels surrounding a
region of interest, the RM values being averaged would be highly correlated
due to convolution with the CLEAN beam, making it very difficult to accurately
estimate the corresponding uncertainty on the average value. For this reason,
all the RM values we have considered are based on single-pixel estimates,
for which the uncertainties are much better understood (Hovatta et al.
2012). Note that this is a conservative approach to estimating the 
statistical significances of the observed RM gradients. 

\subsection{Sign Changes in the Transverse RM Profiles}

The observed Faraday rotation measure depends on both the density of
thermal electrons and the line-of-sight magnetic field in the region
where the Faraday rotation is occurring. Accordingly, gradients in
Faraday rotation could reflect gradients in either the electron density or
line-of-sight magnetic field (or both). However, if an observed gradient is
monotonic and encompasses a change in the sign of the Faraday rotation,
this cannot be explained by an electron-density gradient alone, and 
indicates a systematic change in the line-of-sight magnetic field, such
as would come about in the case of a helical magnetic field geometry in
the region of Faraday rotation. 

Sign changes are observed in the transverse RM gradients detected in
0333+321, 0738+318, 0836+710, 0923+392, 1633+382 and 2345$-$167, 
strengthening the case that these gradients are due to helical
magnetic fields associated with these jets. 

Note that the absence of a sign change in the transverse RM profile
does not rule out the possibility that a transverse gradient is due
to a helical or toroidal field component in the region of Faraday
rotation, since gradients encompassing only one sign can be observed
for some combinations of helical pitch angle and viewing angle.

\subsection{Core-Region Transverse RM Gradients}

In the standard theoretical picture, the VLBI ``core'' represents  
the ``photosphere'' at the base of the jet, where the optical depth is
roughly unity. However, in many cases, the observed core is likely a blend of
this (partially) optically thick region and optically thin regions in 
the innermost jet. Since these optically thin regions are characterized
by  much higher degrees of polarization, they likely dominate the
overall observed ``core'' polarization in many cases. Therefore, we have 
supposed that the polarization angles observed in the core are most likely
orthogonal to the local magnetic field, as expected for predominantly
optically thin regions. We have not observed any sudden jumps in polarization
position angle by roughly $90^{\circ}$ suggesting the presence of optically 
thick--thin transitions in the cores in our frequency range at our epoch.

This picture of the observed VLBI core at centimeter wavelengths corresponding
to a mixture of optically thick and thin regions, with the observed core 
polarization contributed predominantly by optically thin regions, also 
impacts our interpretation of the core-region Faraday rotation measures.
Monotonic transverse RM gradients with significances of at least $3\sigma$
are observed across the core regions of 0333+321, 0738+318, 0923+392, 1150+812,
1633+382, 2037+511 and 2345$-$167. The simplest approach to interpreting 
these gradients is to treat them in the same way as transverse gradients
observed outside the core region, in the jet. While the simulations of 
Broderick et al. (2010) show that relativistic and optical depth effects
can sometimes give rise to non-monotonic transverse RM gradients in core
regions containing helical magnetic fields, there are also cases when these
helical fields give rise to monotonic RM gradients, as they would in a fully
optically thin region. In addition, most of the non-monotonic behaviour
that can arise will be smoothed by convolution with a typical 
centimeter-wavelength VLBA beam [see, for example, the lower right panel in
Fig.~8 of Broderick et al. (2010)]. Therefore, when a smooth, monotonic, 
statistically significant transverse RM gradient is observed across the 
core region, it is reasonable to interpret this as evidence for 
helical/toroidal fields in this region (i.e., in the innermost jet).
 
\subsection{RM-Gradient Reversals}

We have detected distinct regions with transverse Faraday rotation
gradients oriented in opposite directions in 0923+392 and 2037+511.
Similar reversals in the directions of the RM gradients in the core 
region and inner jet have also been reported by Mahmud et al. (2013) for 
the AGNs 0716+714 and 1749+701. 

This at first seems a puzzling result, since the direction of a transverse
RM gradient associated with a helical  magnetic field component
is essentially determined by the direction of rotation of the central
black hole and accretion disk and the initial direction of the poloidal
field component that is ``wound up'' by the rotation. It is difficult in
this simplest picture to imagine how the direction of the resulting
azimuthal field component could change with distance along the jet or
with time. However, in a picture with a nested helical field structure, 
similar to the ``magnetic tower'' model of Lynden--Bell (1996), but
with the direction of the azimuthal field component being different 
in the inner and outer regions of helical field,
such a change in the direction of the net observed RM gradient 
could be due to a change in dominance from the 
inner to the and outer region of helical field in terms of their 
overall contribution to the Faraday rotation (Mahmud et al. 2013).
One theoretical picture that predicts a nested helical-field structure
with oppositely directed azimuthal field components in the inner and
outer regions of helical field is the ``Poynting--Robertson cosmic
battery'' model of Contopoulos et al. (2009) [see also references
therein], although other plausible systems of currents may also
give rise to similar field configurations.

\section{Conclusion}

We have presented new polarization and Faraday RM measurements of eight
AGNs based on 4.6--15.4~GHz observations with the Very Long Baseline
Array. These objects were part of a sample of sources displaying evidence
for polarization structures indicating a ``spine'' of transverse magnetic
field flanked by a ``sheath'' of longitudinal magnetic field on one or
both sides of the jet at one or more epochs in the 15.4-GHz MOJAVE images.
One possible origin for this type of polarization structure is a helical
jet magnetic field: depending on the pitch angle of the helical field and
the viewing angle of the jet, the sky projection of the field may be
predominantly orthogonal to the jet near the central ridge line, becoming
predominantly longitudinal near the jet edges. If these jets do carry
helical magnetic fields, their Faraday rotation measure (RM) distributions
should also show a tendency to display transverse RM gradients, whose origin
is the systematic change in the line-of-sight component of the helical
field across the jet.

All eight of these AGNs show evidence for Faraday rotation measure (RM) 
gradients across their core regions and/or jets with statistical
significances of at least $3\sigma$. Two more AGNs with spine--sheath
polarization strucure observed in the same
VLBA experiment showed no signs of such systematic transverse RM gradients.
The fact that eight of ten AGNs displaying partial or full 
``spine--sheath'' transverse polarization structures also show firm 
evidence for transverse RM gradients supports the hypothesis that both 
of these properties are due to the presence of a helical magnetic 
field associated with the VLBI jets of these AGNs. It is also striking
that sign changes in the transverse RM profiles are observed for six of
these eight AGNs, which can only be explained by a gradient in the
line-of-sight magnetic field, not a gradient in electron density.

Two of the eight AGNs considered here  --- 0923+392 and 2037+511 ---
also show evidence for distinct regions with transverse Faraday rotation
gradients oriented in opposite directions. This structure can also be
seen in the RM map of 0923+392 presented by Zavala \& Taylor (2003).
Similar reversals in the directions of the RM gradients in the core
region and inner jet have been reported by Mahmud et al. (2013) for
the AGNs 0716+714 and 1749+701. This seemingly puzzling result can be
explained in a picture with a nested helical field structure, with 
the direction of the azimuthal field component being different in the 
inner and outer regions of helical field. In this case, the reversal
of the transverse RM gradient is due to a change in dominance from the
inner to the and outer region of helical field in terms of their
overall contribution to the Faraday rotation (Mahmud et al. 2013).

\section{Acknowledgements}

We thank Sebastian Knuettel for help in the preparation of this 
manuscript, Robert Zavala for providing the version of the AIPS
task RM used for this work, and Gregory Tsarevsky for
compiling his online catalog of integrated Faraday rotation measures,
which was helpful in our analysis 
(http://lukash.asc.rssi.ru/people/Gregory.Tsarevsky/RM Cat96.html). 
We are also grateful to the anonymous referee for his or her quick
response and thoughtful comments.
This research has made use of data from 
the University of Michigan Radio Astronomy Observatory which has been 
supported by the University of Michigan and by a series of grants from 
the National Science Foundation, most recently AST-0607523.

\end{document}